\documentstyle[multicol,prl,aps,epsf,psfig]{revtex}

\begin {document}

\title
{
Self-Organization in a Granular Medium by Internal Avalanches
}
\author
{
S. S. Manna
}
\address
{
Satyendra Nath Bose National Centre for Basic Sciences,
    Block-JD, Sector-III, Salt Lake, Calcutta-700098, India \\
}
\maketitle
\begin{abstract}

Internal avalanches of grain displacements can be created inside a 
granular material kept in a bin in two ways: (i) By 
removing a radomly selected grain at the bottom of the bin
(ii) By breaking a stable arch of grains clogging a hole at the bottom
of the bin. Repeated generations of such avalanches lead the
system to a steady state. The question asked, is this state a critical
state as that in Self-Organized Criticality? 
We review here some of the recent studies on this problem using
cellular automata as well as hard disc models.
\vskip 0.3 cm
PACS number(s): 05.70.Jk Critical Point Phenomena - 74.80.Bj Granular, 
melt-textured, and amorphous superconductors; powders

\end{abstract}

\begin{multicols}{2}

\vskip 0.2 cm
\leftline {\bf 1. Introduction}
\vskip 0.2 cm

A granular material is a collection of a large number of solid grains,
such as sand, seeds, sugar, salt, stone chips etc. \cite {hansen,grain}. A granular
material can be poured like a liquid into a container and also can form
a pile against gravity like a solid. For this reason, a granular material
is widely regarded as the fourth state of matter. 

Use of granular materials is quite extensive in the industry and in the
daily lives. Powders are routinely used in the food and pharmaceutical
industries.  Large quantities of sand and stone chips are used for the
construction of highways and barrages. Food materials like flour, cereals,
carbon, salts are transported, processed and stored. Apart from these the
geological processes like landslides and snow avalanches on mountains
involve granular materials. In spite of so many applications, flow
of grains in granular materials and packing phenomena are not well
understood from a fundamental viewpoint though the study of granular
materials goes back to few centuries \cite {coulomb}.

Recently a new idea was proposed by Bak, Tang and Wiesenfeld (BTW) 
to explain the formation of avalanches on the surface of a sandpile
and is called `Self-Organized Criticality' (SOC) \cite {btw}.
Self-Organized Criticality is the emergence of long-ranged spatio-temporal
correlations in non-equilibrium steady states of slowly driven systems 
without fine tuning of any control parameter.
It says that there are certain class of systems in nature which become critical
under their own dynamical evolutions. An external agent triggers into a system a
transport process of some physical entity, say mass. The transport 
process is made of a large number of microscopic relaxation events.
This cascade of events together is called an avalanche. After largre 
number of avalanches, the
system arrives at a {\it critical state}, when the avalanches extend over all length and
time scales \cite {btw,soc,socbook}.

  BTW observed that a typical sandpile satisfies all the criteria of SOC.
Consider a sandpile formed on a fixed horizontal base, with an
arbitrary initial distribution of sand. The pile is grown by adding repeatedly
a few grains of sand at a time. After a long time the sandpile takes a
fixed conical shape and the system is said to reach the steady state.
Addition of each sand grain results some
activity on the surface of the pile: an avalanche of sand mass followed,
which propagates on the surface of the sandpile. These avalanches are of
many different sizes and BTW expected that they should have a power law
distribution in the steady state.

  Laboratory experiments on sand piles, however, have not been fully successful
yet to show unambiguously the existence of SOC in sandpile systems. In the
first experiment, the granular material was kept in a semicircular drum which
was slowly rotated about the horizontal axis, thus slowly tilting the free surface
of the pile \cite{chicago}. Grains fell vertically downward and were allowed to pass
through the plates of a capacitor. Power spectrum analysis of the time series
for the fluctuating capacitance however showed a broad peak, contrary to the
expectation of a power law decay, from the ideas of SOC \cite{chicago}.

  In a second experiment, sand mass was slowly dropped on to a horizontal
circular disc, to form a conical pile in the steady state \cite{ibm}. On further
sand addition, sand avalanches were created on the surface of the pile, and
the outflow statistics was observed. Size of the avalanche was measured
by the amount of sand mass that dropped out of the system. It was observed
that the avalanche size distribution obeys a scaling behaviour for small
sizes of the base. For large base, however the scaling did not work very well.
It was suggested that the observed SOC in this system is a finite size effect \cite{ibm}.
However, it was claimed that a stretched exponential distribution for the avalanches
fits the whole range of system sizes \cite {feder}.
The reasons for not observing scaling are the existence of two angles of repose
as observed in \cite {chicago} and also the effect of inertia of
the sand mass while moving down during avalanches. 

  This effect was suppressed in an experiment with anisotropic rice grains 
which used a pile of rice between two vertical glass plates
separated by a small gap \cite{oslo}. Rice grains were slowly dropped on to the pile.
Due to the anisotropy of grains, various packing configurations were observed.
In the steady state, avalanches of moving rice grains refreshed the surface
repeatedly. SOC behaviour was observed for grains of large aspect ratio, but
not for the less elongated grains \cite{oslo}.

\vskip 0.2 cm
\leftline {\bf 2. Internal Avalanches in a Granular Medium}
\vskip 0.2 cm

   In all the studies discussed above the avalanches propagate on the
surfaces of the sandpiles. However, there exists the possibility of
creating avalanches in the interior of a granular material. In a
granular material kept in a bin at rest, different grains support
one another by mutually acting balanced forces. Now if a grain is
removed, the grains which were supported by it become unstable and
tend to move. Eventually the grains in the further neighbourhood
also loose their stability. As a result an avalanche of grain
displacements takes place, which results the upward
propagation of some void space. The activity stops when no more grains
remain unstable. We call this avalanche as the {\it Internal Avalanche}.

\vskip -5.0 cm
\begin{figure}
\begin{center}
\centerline{\epsfxsize=10.96 cm{\epsfysize=15.0 cm\epsfbox{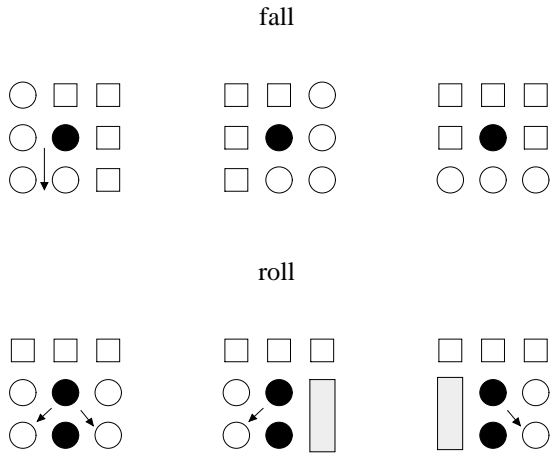}}}
\vskip -2.5 cm
\narrowtext{\caption{
The possible $fall$ and $roll$ moves in the cellular
automata model of the granular system on the square lattice.
Filled circle denotes the position of a grain, unfilled circle
denotes a vacant site. The grain moves to the vacant position
irrespective of the occupation of the sites with square boxes.
Shaded rectangle denotes a pair of sites, in which at least
one is occupied.
}}
\end{center}
\end{figure}

  Depending on the method of perturbing the system we consider
two kinds of internal avalanches. In its first kind, a randomly
selected grain at the bottom of the granular bin is removed. 
The other way of creating avalanches is to make a hole at the centre
of the bin and allow some granular mass to flow out. This flow
soon stops by the formation of an arch which cloggs the hole.
The arch is then broken by removing one of its grains.
We call these as internal avalanches of the second kind.

  We believe, the internal avalanches are 
better candidates for observing SOC. Due to the high compactness 
of grains in the granular material in a bin, a grain never gets 
sufficient time to accelerate much and therefore the effect of 
inertia should be small. The basic physical behavior here is thus quite
different from the avalanches on the surface of the pile because
of the constraints to particle motion in the dense particle beds.

\vskip -6.5 cm
\begin{figure}
\hspace*{0.6 cm}
\centerline{\epsfxsize=12.0 cm{\epsfysize=15.0 cm\epsfbox{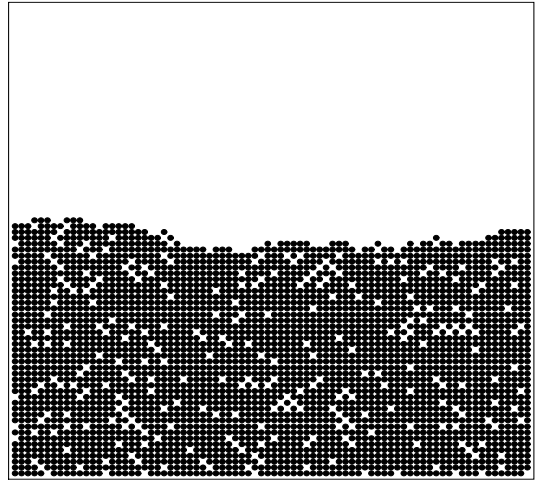}}}
\vskip -1.8 cm
\narrowtext{\caption{
The initial configuration grains in the Cellular Automata model. The 
bin of size $80 \times 80$ contains 2976 grains.
}}
\end{figure}

   Internal avalanches were first studied in a two-dimensional semi-lattice 
model \cite{snyder}. Non-overlaping unit square blocks model the grains, whose
horizontal coordinates can vary continuously where as the vertical
co-ordinates are discretized. A grain can only fall vertically
if insufficiently supported and sufficient space below is available.
The system is disturbed by repeatedly removing grains one after another
at the bottom of the bin and thus creating avalanches of grain movements.
While this model and also other cellular automata \cite {disc} and 
`Tetris-like' models \cite {tetris} did exhibit scaling behaviour for 
the internal avalanches, the disc models \cite {disc,bang,manna} 
in continuous space, however, did not show sufficient evidence of SOC.

\vskip 0.2 cm
\leftline {\bf 3. Cellular Automata Model}
\vskip 0.2 cm

  In the cellular automata model of a granular system an occupied 
(vacant) site on a lattice represents the presence (absence) of a
grain \cite{disc}. A square lattice of size $L$ with periodic boundary condition 
along the $x$ axis represents the bin. The gravity acts in the $-y$ 
direction. The initial configuration of grains in the bin is 
generated by the ballistic deposition method. Grains are dropped
along the randomly chosen vertical directions. After landing on the
surface of the granular heap, the grain relaxes to its lowest 
energy position on the surface.

  The different grain moves are explained in Fig. 1. A gain can have
two types of possible moves in unit time: It can {\it fall} a lattice
unit vertically, or it can {\it roll} either to the lower left or to the
lower right neighbouring positions. A single movement of a grain at 
$C(i,j)$ involves the neighbouring seven sites: $LU(i-1,j+1), L(i-1,j), 
LD(i-1,j-1), D(i,j-1), RD(i+1,j-1), R(i+1,j)$ and $RU(i+1,j+1)$.
In the $fall$ move the grain comes down one level to the vacant
site at $D$ and in $roll$ move the grain goes either to the vacant 
site at $LD$ or at $RD$ (Fig. 1). An arch is formed when a grain 
at $C$ is considered stable if both of its two diagonally opposite 
sites either at $LD$ and $RU$ or, at $LU$ and $RD$ are occupied.
As a result, on the square lattice, the only possible shape of an arch
consists of two sides of a triangle. 
Initial granular patterns generated using the random ballistic
deposition method with restructuring (BDRM) \cite{meakin} and patterns are same for both
models (Fig. 2). Depending on whether we allow the
arch formations or not, we define the following two models.

\vskip -6.5 cm
\begin{figure}
\hspace*{0.6 cm}
\centerline{\epsfxsize=12.0 cm{\epsfysize=15.0 cm\epsfbox{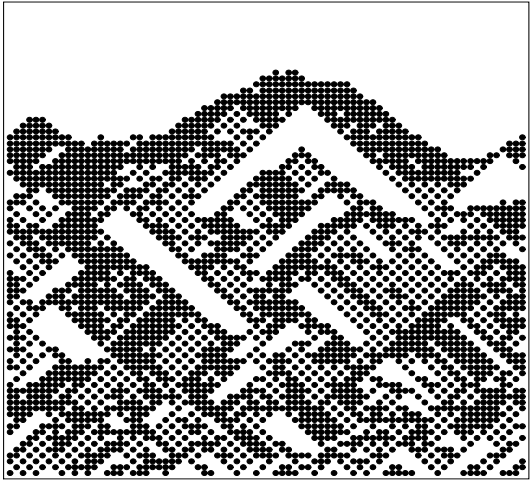}}}
\vskip -1.8 cm
\narrowtext{\caption{
A steady state granular pattern for the model A after a large number
of avalanches starting from the initial pattern as in Fig. 2.
}}
\end{figure}

   In model A we allow arch formations. The grain at $C$ falls only 
if any of the following three conditions is satisfied:
(i)   $LD$, $L$ and $LU$ are vacant
(ii)  $RD$, $R$ and $RU$ are vacant
(iii) $LD$ and $RD$ are vacant.
In all other situations the grain does not fall. Notice that in conditions (i)
and (ii) we are allowing the formation of arches.
The grain at $C$ rolls only if the site $D$ is occupied. This is done in any of
the three following ways:
(i) If both $LD,L$ are vacant but either of $RD,R$ is occupied then the grain
rolls to $LD$.
(ii) if both $RD,R$ are vacant but either of $LD,L$ is occupied then the grain
rolls to $RD$.
(iii) If all four sites at $LD, L, RD, R$ are vacant then the grain rolls
either to $LD$ or to $RD$ with probability $1/2$. A steady state pattern 
for the model A after a large number of avalanches is shown in Fig. 3.

   In model B we do not allow arch formations. The first two conditions for
fall of model $A$ are modified as:
(i)  $LD$ and $L$ are vacant
(ii) $RD$ and $R$ are vacant. All other conditions of fall as well as roll
remain same as in model A. A steady state pattern for the model B after a 
large number of avalanches is shown in Fig. 4.

\vskip -6.0 cm
\begin{figure}
\hspace*{0.6 cm}
\centerline{\epsfxsize=12.0 cm{\epsfysize=15.0 cm\epsfbox{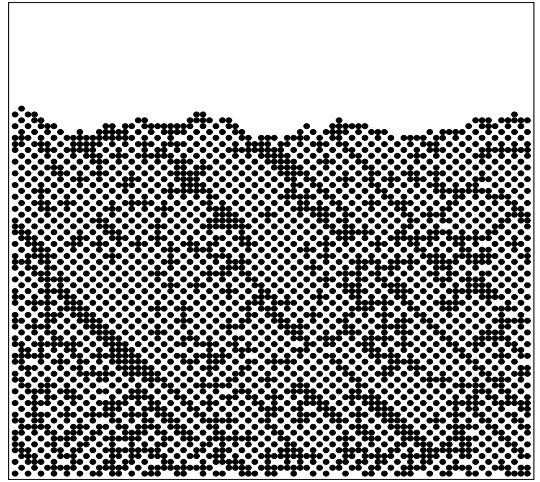}}}
\vskip -1.8 cm
\narrowtext{\caption{
A steady state granular pattern for the model B after a large number
of avalanches starting from the initial pattern as in Fig. 2.
}}
\end{figure}

\vskip -4.0 cm
\begin{figure}
\hspace*{0.3 cm}
\centerline{\epsfxsize=9.0 cm{\epsfysize=12.0 cm\epsfbox{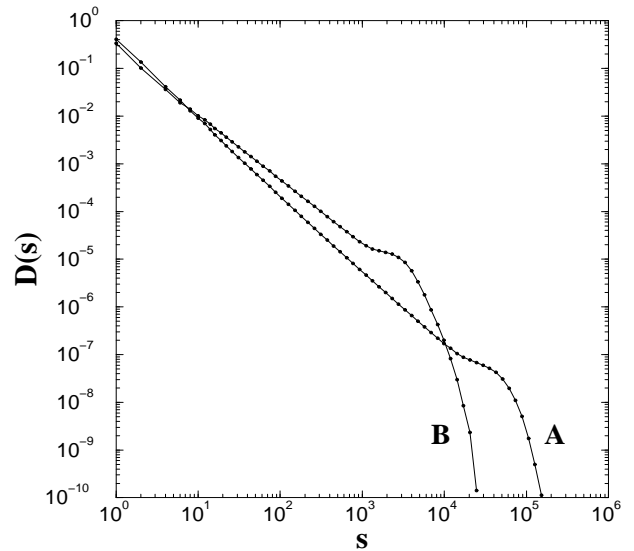}}}
\vskip -.8 cm
\narrowtext{\caption{
The coarse grained distributions of avalanche sizes for the models A and B 
for system sizes $256 \times 256$.
}}
\end{figure}

  The avalanches are created by taking out one grain at a
time at the bottom, allowing the system to relax and replacing it randomly
on the surface after the avalanche is over. We see that the average
density of sites, starting from
an initial value of $0.907 \pm 0.005$, decreases to the final stable
value of $0.590 \pm 0.005$ in model A and to $0.618 \pm 0.005$ in model B.
The avalanche size $s$ and life time $t$ follow power law distributions:
$D(s) \sim s^{-\tau^A_s}, D(t) \sim t^{-\tau^A_t}$ and similarly for the
model B (Fig. 5). Different exponents are obtained for the two models:
$\tau^A_s \approx 1.48$  and $\tau^A_t \approx 1.99$ where as
$\tau^B_s \approx 1.34$ and $\tau^B_t \approx 1.50$. 

\vskip 0.2 cm
\leftline {\bf 4. The Tetris Model}
\vskip 0.2 cm

  The Tetris model is defined on an oriented (at 45$^o$ with the $y$ axis)
square lattice with periodic boundary condition along the $x$ direction,
gravity acts along the $-y$ axis \cite {tetris,supriya1,supriya2}.
The bottom is closed and grains are released at the top. The grains in this
model are of asymmetric shapes. In the simplest version of the model,
particles are represented by rectangular plaquettes of sizes $a \times b$.
These grains are positioned at the lattice sites with two possible
orientations i.e., the length of the grains can be aligned along any of the
two principal axes of the lattice. The size of the grain is chosen in lattice units as:
$a > 1/2$ and $a+b < 1$. Excluded volume effect is strictly maintained,
that is overlaping of any two grains is just not allowed. An immediate
consequence of this result is no two grains with their axes along the
same principal axis can be positioned at the neighbouring lattice sites,
however, they can position at neighbouring sites with their axes parallel.

\begin{figure}
\begin{center}
\centerline{\epsfxsize=8.0 cm{\epsfysize=8.0 cm\epsfbox{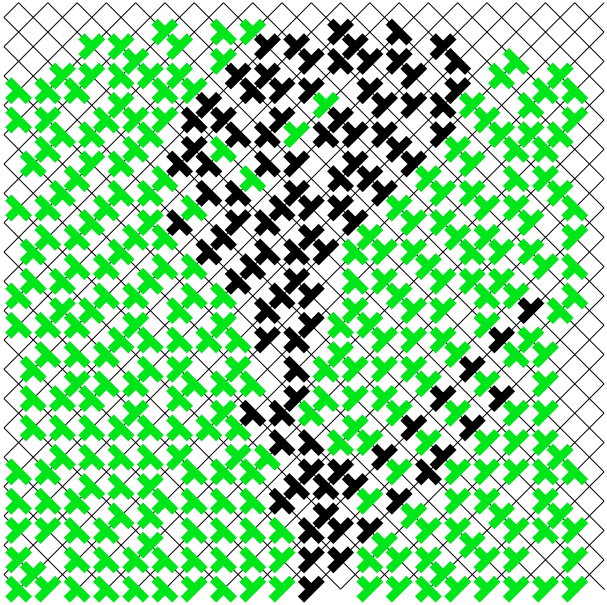}}}
\vskip 0.5 cm
\narrowtext{\caption{
A steady state configuration of the Tetris model
where anisotropic grains have the shape of a T.
Grains can have two different orientations.
The boundary conditions are periodic
in the horizontal direction. The black grains are disturbed   
in the avalanche caused by removing the lowest black grain.
}}
\end{center}
\end{figure}

  The initial stack of grains in the bin of size $L_x \times L_y$ is
generated by releasing grains one after the other at the top 
level of $y=L_y$. The orientation of a grain along either of the two
lattice axes is randomly selected with equal probability. 
Grains retain their orientations as they move and are not allowed to
rotate. The grain
then executes a directed random walk, at each step it makes a jump
to one of the two downward neighbours which are along and perpendicular
to its axis. It settles at a site from which it cannot come down further
due to the excluded volume effect \cite{supriya1,supriya2}.

  The granular system is then repeatedly driven by removing one grain at
a time from the bottom. Each time a grain is removed, it makes the
neighbouring grains unstable. This instability propagates upwards 
involving other grains in the system in the form of an internal avalanche
untill it gets arrested, when no further grains move and the system 
becomes stable again. The grain is the dropped again from the top
to conserve grain number in the system.
 
  The size of the avalanche is defined as the total number of grains
destabilized by the process of removing a grain at the bottom. The
size distribution of the avalanches follows a power law: $D(s) \sim s^{-\tau}$.
Numerical simulations suggests a value for $\tau=1.5 \pm 0.05$ \cite{supriya1,supriya2}.

\vskip -12.5 cm
\begin{figure}
\begin{center}
\hspace*{-6.5 cm}
\centerline{\epsfxsize=15.0 cm{\epsfysize=20.0 cm\epsfbox{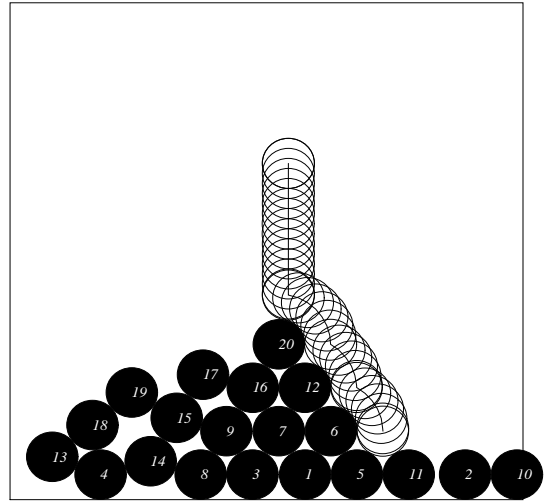}}}
\vskip 0.5 cm
\narrowtext{\caption{
Procedure for adding a single grain to the surface of
a granular heap in a bin, using the Ballistic deposition by restructuring method.
The already formed heap has 20 grains. The 21-st grain is being deposited
in a sequence of successive `fall' and `roll' moves.
}}
\end{center}
\end{figure}

\vskip 0.2 cm
\leftline {\bf 5. The Hard Disc Model}
\vskip 0.2 cm

   During the propagation of an avalanche in a granular medium
grains compete locally with one another to occupy the same vacant space.
The high packing densities of the grains in the  bed prevent a single grain
from occupying the available void space, and consequently grains
get locked to form `arches' \cite{grain}. A stable arch in a two dimensional
granular pattern is a chain of grains
where the weight of each grain is balanced by the reaction forces from
two neighbouring grains in the chain. Arches can form only when a grain
is allowed to roll over other grains. 

\vskip -8.0 cm
\begin{figure}
\begin{center}
\hspace*{1.0 cm}
\centerline{\epsfxsize=13.0 cm{\epsfysize=18.0 cm\epsfbox{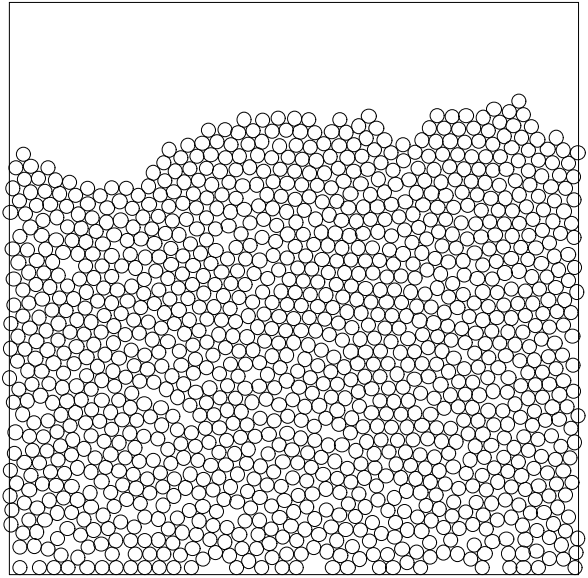}}}
\vskip -2.0 cm
\narrowtext{\caption{
A stable state configuration of the hard disc model after a large number
of avalanches in a bin of size $80 \times 80$ containing 1200 grains.
}}
\end{center}
\end{figure}

   The arch formations are very well reproduced if the grains are modeled
as hard spheres. We study the two dimensional version of this system. 
The granular material is a collection of grains modeled by $N$ hard
mono-disperse discs of radii $r$. No two grains are allowed to overlap
but can come very close to each other. In fact a grain can touch another
grain and roll on it. The granular bin is 
represented by a rectangular area on the $x-y$ plane: from $x=0$ to $L_x$ 
and $y=0$ to $L_y$. Periodic boundary condition is imposed along the $x$ 
direction and gravity acts along the $-y$ direction. The bottom of the bin 
coinciding with $y=0$ line is highly sticky and any grain which comes 
in its contact gets stuck there and does not move further.

   The initial grain pattern is generated by the `ballistic deposition and
restructuring method (BDRM)'\cite{meakin} (Fig. 7). In this method, grains are released at random 
positions at the top level of $y=L_y$ sequentially one after the other. 
Subsequently, they are allowed to fall vertically till they come in contact 
with the pile when they roll down to their stable positions along the paths 
of steepest descent. When all $N$ grains are dropped, we have the initial 
configuration (Fig. 8). 

   Each grain is assigned a serial number $n$ from 1 to $N$.The centre coordinates 
of these grains are stored in the $x_N(n)$ and $y_N(n)$ arrays. Very often 
we search the set of grains that reside in the local
neighbourhood of a particular grain $n$. As a brute force solution
one can calculate the distances $r_{nm}$ of the centres of all other
grains $(m=1,N; m \ne n)$ from the centre of the $n$-th grain and
pick up those grains which are within the local neighbourhood.
This takes CPU proportional to $N$.

\begin{figure}
\begin{center}
\epsfxsize=3.3in
\epsffile{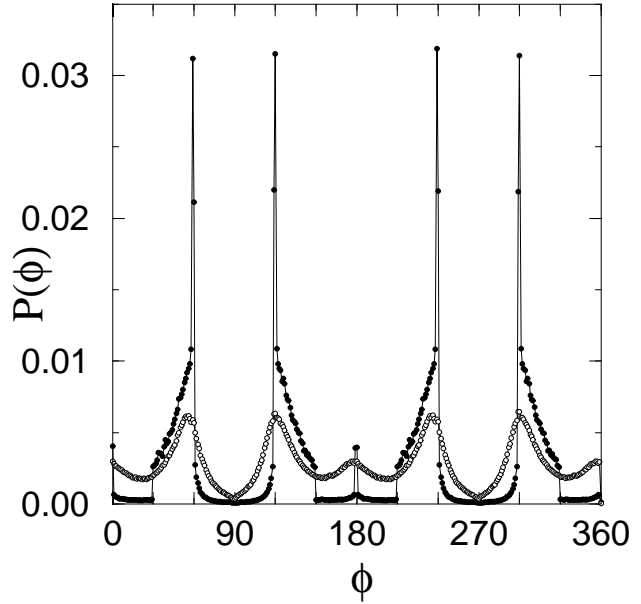}
\narrowtext{\caption{
Distribution $P(\phi)$ of the angles $\phi$ of the contact vectors.
The plot for the initial distribution is shown by black
dots. It has four high peaks at $60^o, 120^o, 240^o, 300^o$. The curve with
opaque circles corresponds to the steady state. The randomness introduced
into the system is reflected by the drastic reduction of the height of the
peaks as well as a broadening.
}}
\end{center}
\end{figure}
\vskip -1.0 cm

   The search is done more efficiently by considering an underlying grid
$S$. A primitive cell is referred by the coordinates of its bottom-left
corner $(i,j)$. If the centre of a grain $n$ with coordinates $(x_n,y_n)$
is within the primitive cell $(i,j)$ then the location $(i,j)$ in the $S$
array is assigned the grain number $n$. A primitive cell can contain the 
centre of at most one grain if the radii of the grains are chosen as 
$r= 1/\sqrt 2 + \epsilon$, where $\epsilon$ is a small 
positive number.  Therefore if the $S(i,j)$ location is zero, it implies 
that the cell $(i,j)$ does not contain the centre of any grain.

  The local neighbourhood $LN$ of a grain whose centre is within
the cell $(i,j)$ by the area extending from $(x=i-2 $ to $ x < i+3$ and
$y=j-2 $ to $ y < j+3)$. Only the grains, whose centres are
within $LN$ may overlap with the grain whose centre is within the cell $(i,j)$.
While searching for the set of grains within $LN$, we search the lattice locations
$i-2$ to $i+2$ and $j-2$ to $j+2$. Numbers stored in these locations give
the grain numbers and their centre coordinates are obtained from the $x_N(n)$
and $y_N(n)$ arrays. This search takes CPU proportional to $N^0$.

   The temporal evolution of the granular system is studied by a
`pseudo-dynamics'. Unlike the method of molecular dynamics we do not
solve here the classical equations of motion for the grain system.
Only the direction of gravity and the local geometrical constrains
due to the presence of other grains govern the movement of a grain.
In our method a grain can have only two possible movements namely
the un-obstructed vertical $fall$ and the un-obstructed $roll$ over another grain
in contact. A parameter $\delta$ is introduced to characterize the
fall and the roll moves. Time is discretized and in a unit time a grain
may fall to a maximum height of $\delta$ or its centre can roll a
maximum angle of $\theta=\delta/2r$ over another grain. During both
fall and roll a grain does not accelerate. We believe this dynamics models
an assembly of very light grains in 0+ gravity.

\begin{figure}
\begin{center}
\epsfxsize=3.3in
\epsffile{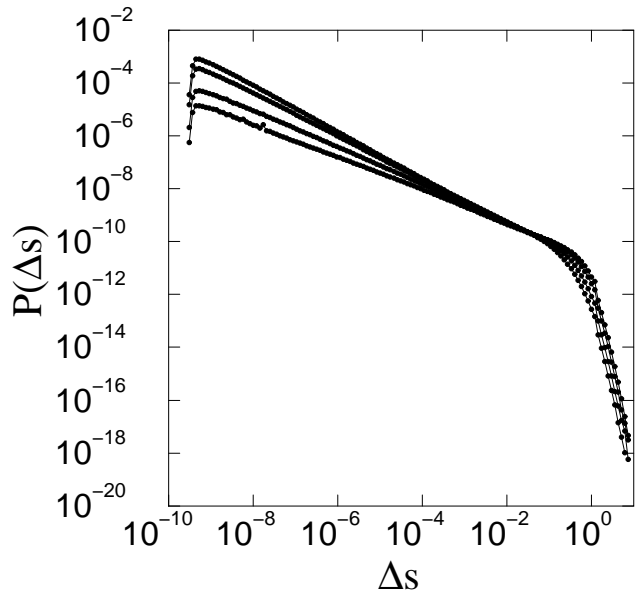}
\narrowtext{\caption{
Distribution $P(\Delta s)$ of the displacement $\Delta s$ of the grains
in a granular system.
Curves for different sizes with grain numbers $N$ = 300, 625, 2500 and
10000 are plotted from bottom to top on the left.
}}
\end{center}
\end{figure}
\vskip -0.8 cm

   Since the grains are all of equal radii, a grain may be in contact
with a maximum of six other grains. To recall quickly the set of grains
in contact with an arbitrarily selected grain $n$ we store the serial numbers
of the contact grains for every grain into an array
$neb(N,6)$. The grain $n$ may be at rest on two other grains and its
weight is balanced by the reaction forces from them. The centres of
the supporting grains must be on the two opposite sites of the vertical
line passing through the centre of the stable grain $n$. We reserve the
location $neb(n,1)$ to store the serial number $n_L$ of the left supporting
grain and $neb(n,2)$ to store the serial number $n_R$ of the right
supporting grain. The serial numbers of all other grains which are in
contact with the grain $n$ are kept one after the other in any order
in the locations $neb(n,m),m=3,6$.

   To find out the neighbours in contact with a grain $n$ we follow the
following procedure. We first make a list of grains whose centres are within
the local neighbourhood $LN$ of the grain $n$. From this list, we sort out
those grains whose centres are within a distance $2r-\epsilon_1$ to $2r+\epsilon_1$
from the centre of $n$, $\epsilon_1$ being the tolerance factor which takes
care of the accumulation of errors originated from real number manipulations. We consider
these grains are in contact to the grain $n$. For each contact grain we
calculate the angle $\psi$ measured from the vertically downward direction
through $(x_n,y_n)$ to the vector starting from $(x_n,y_n)$ to the centre
of the contact grain. This angle $\psi$ is measured $+$ve in the anti-clockwise
direction and $-$ve in the clockwise direction. We first find out two contact
grains with minimum values of $\psi$ in the positive and the negative
directions. If the sum of the magnitudes of these two angles is less than
$\pi$, the grain $n$ is considered at rest.

   When a grain $n$ is updated, it first selects the type of movement it is
going to take:

\begin {itemize}

\item [{\bf a. }] If $n_L = n_R = 0$ the grain $n$ is allowed to fall.

\item [{\bf b. }] If $n_L \ne 0$ but $n_R = 0$ the grain $n$ is allowed
to roll on the right over the supporting grain $n_L$ in contact at the left.

\item [{\bf c. }] If $n_L = 0$ but $n_R \ne 0$ the grain $n$ is allowed
to roll on the left over the supporting grain $n_R$ in contact at the right.

\item [{\bf d. }] If $n_L \ne 0$ and $n_R \ne 0$ the grain $n$ is
considered a stable and does not move.

\end {itemize}

    The $fall$ movement of a grain $n$ is executed in the following way.
Corresponding to every grain in the local neighbourhood $LN$ we first
calculate the distance through which $n$ should come down vertically to
make a contact. The minimum $\delta_m$ of these distances corresponds
to the grain $t$ with the centre at $(x_t,y_t)$. If $\delta_m < \delta$
the grain $n$ is brought down a distance $\delta_m$ so that it just
touches $t$. The new coordinates are given by:
\[
x_n'=x_n, \quad \quad y_n'=y_t+\sqrt{4r^2-(x_n-x_t)^2}.
\]
However if $\delta_m \ge \delta$, the grain $n$ falls a
distance $\delta$ only.
The neighbour lists of the grains $n$ and $t$ are updated.
The lattice $S$ is updated for the movement
of the grain $n$ and therefore the lattice point corresponding to $(x_n,y_n)$
is vacated whereas that corresponding to $(x_n',y_n')$ is occupied.

   Now consider the situation when a grain $n$ $rolls$ over another grain $t$
with the centre at $(x_t,y_t)$. The minimum angle $\theta_m$ through which
the grain $n$ should freely roll to be in touch with another grain $m$
in the neighbourhood $LN$ with the centre at $(x_m,y_m)$ is calculated.
If $\theta_m < \theta$ the grain $n$ is rolled an angle $\theta_m$
over the grain $t$ so that it simultaneously touches both the grains
$t$ and $m$. The centre coordinates $(x_n',y_n')$ of the grain $n$ in the new
position are given by:
\[
x_n' = \frac{1}{2}(x_m+x_t) \pm
(y_m-y_t)\sqrt{\frac{4r^2}{d_{mt}^2}-\frac{1}{4}}
\]
\[
y_n' = \frac{1}{2}(y_m+y_t) \mp
(x_m-x_t)\sqrt{\frac{4r^2}{d_{mt}^2}-\frac{1}{4}}.
\]
where $d_{mt}$ is the distance between the centres of the grains $m$ and $t$.
If the angle $\theta_m \ge \theta$, the grain $n$ rolls an angle $\theta$
over the grain $t$. In both cases if it happens that $y_n' < y_n$, then
the grain $n$ is brought at the same level as the grain $t$ and 
$x_n'=x_n \pm 2r$ and $y_n'=y_n$ are assigned. The lattice $S$ is renewed for the movement
of the grain $n$ and therefore the lattice point corresponding to $(x_n,y_n)$
is vacated whereas that corresponding to $(x_n',y_n')$ is occupied.
The $\pm$ signs are for the left and right rolls.
The neighbour lists of the grains $n$ and $t$ are also
updated.

\vskip -2.7 cm
\begin{figure}
\begin{center}
\hspace*{-0.5cm}
\centerline{\epsfxsize=9.2 cm{\epsfysize=12.5 cm\epsfbox{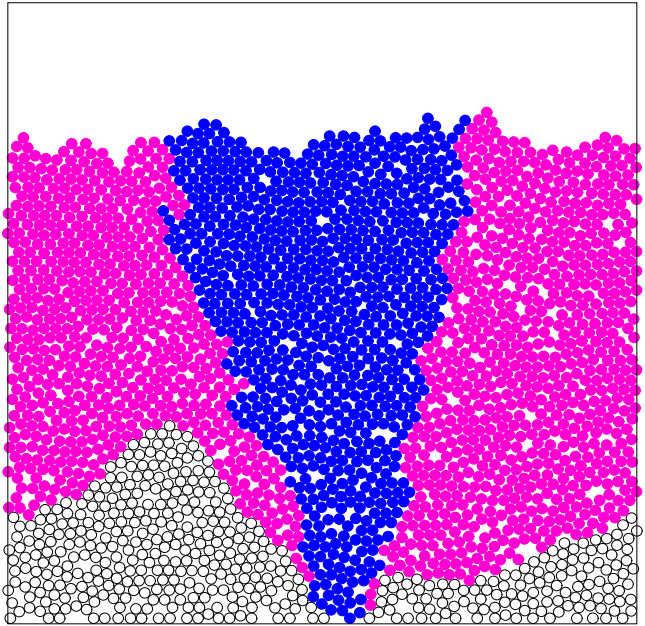}}}
\vskip -0.7 cm
\narrowtext{\caption{
The steady state configuration for a system of 2500 grains
in a bin of $L_x=80$. The gray and the black circles denote the
grains in the supporting cluster of the lower most circle at the
bottom. If this circle is deleted, only the black circles will
move, which constitute the avalanche cluster. The vacant circles
are the undisturbed grains.
}}
\end{center}
\end{figure}
\vskip -0.5 cm

\vskip 0.2 cm
\leftline {\bf 5.1. Internal Avalanches of the First Kind}
\vskip 0.2 cm

   The system of $N$ grains resting in the bin are disturbed by
removing an arbitrary grain $n$ at the bottom of the bin.
A list of the discs residing at the bottom is made. To create an avalanche,
a member of the list is randomly chosen and the corresponding   
disc at the bottom is deleted at time $t$=0.
A number of neighbouring grains
depend on the grain $n$ for their stability. These grains will be unstable
and start to move and so also their further neighbours. 
A grain $n$ is updated by checking its first two locations in the $neb$ array i.e.,
$neb(n,1)$ and $neb(n,2)$ and determine which type of movement it
is going to execute. In general, at any time $t$ we have a list of unstable grains.
When we update them some become stable but others remain unstable. The
new list at time $t+1$ consists of these unstable discs from time
$t$ in addition to the neighbouring grains which become unstable due to motion of grains in time
$t$. The avalanche of grain displacement stops when the list is empty.
In a certain time $t$ the grains are updated sequentially.
After the avalanche stops, the removed grain is kept back at the
surface of the pile by releasing again at an arbitrary position at the top
level.


\centerline{\bf (a)}

\vskip -7.8 cm
\begin{figure}
\hspace*{2.0cm}
\centerline{\epsfxsize=12.0 cm{\epsfysize=15.0 cm\epsfbox{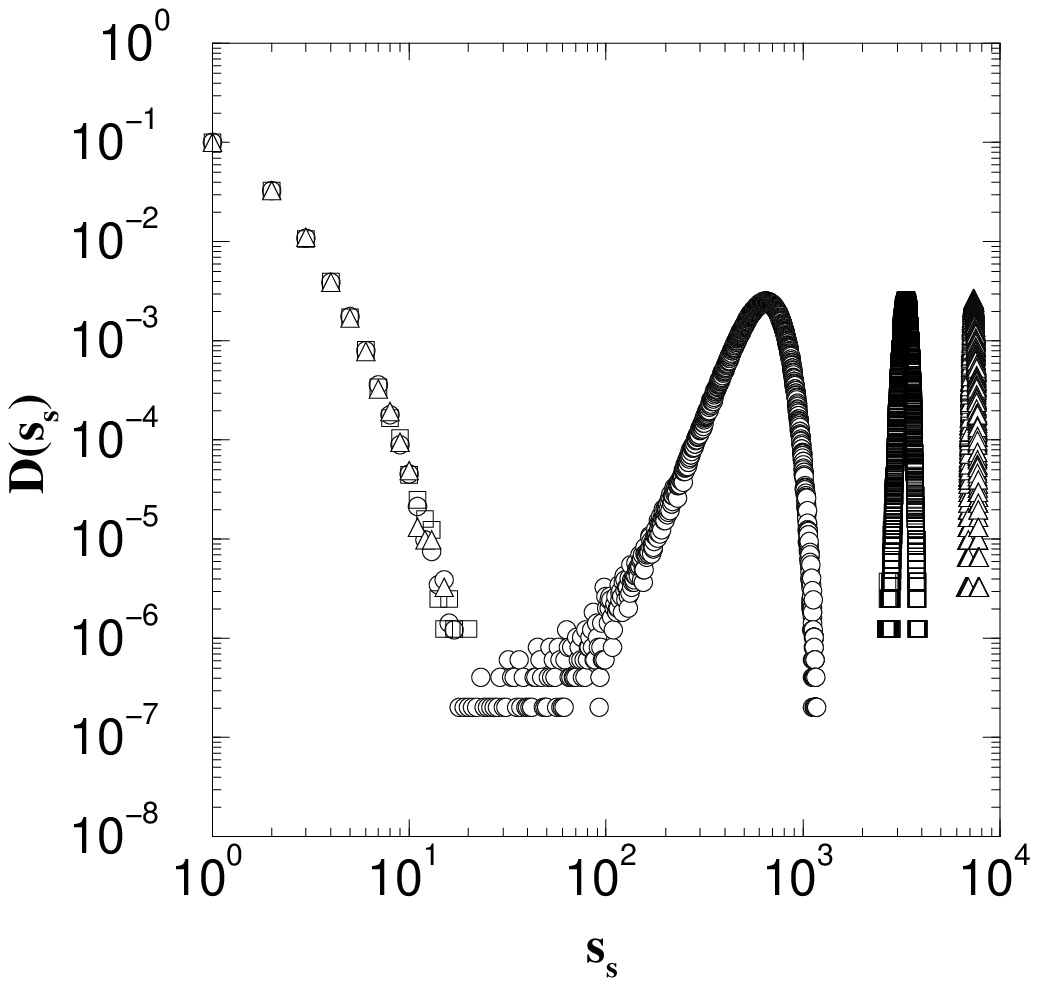}}}
\end{figure}

\vskip -1.5 cm
\centerline{\bf (b)}

\vskip -7.5 cm
\begin{figure}
\hspace*{2.0cm}
\centerline{\epsfxsize=12.0 cm{\epsfysize=15.0 cm\epsfbox{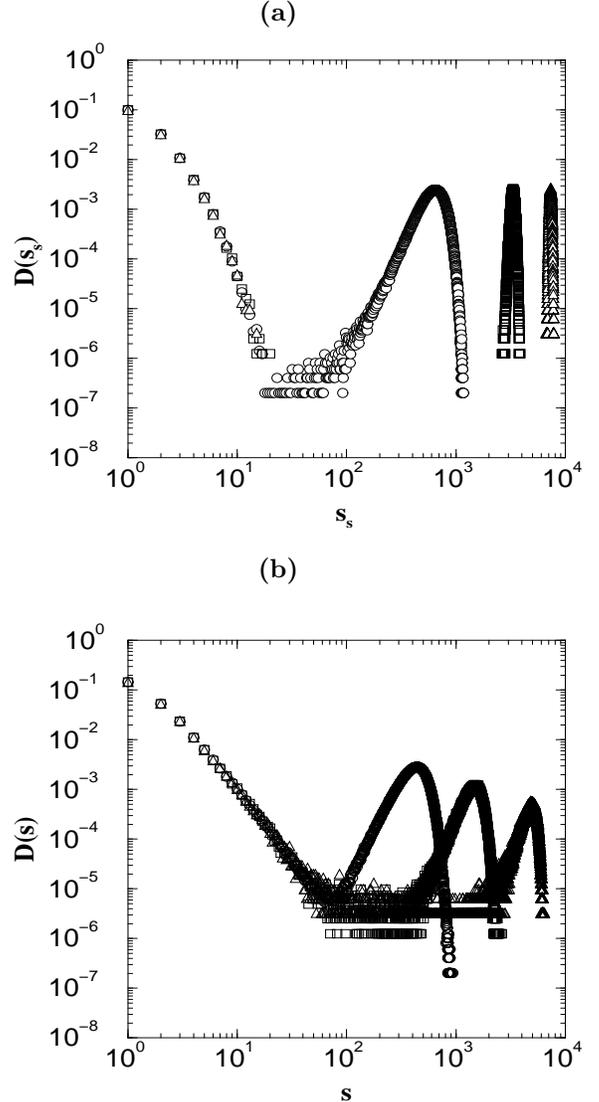}}}
\vskip -0.7 cm
\narrowtext{\caption{
(a) Distributions of the  supporting clusters $(s_s)$
for the systems of 2000 (circle), 4000 (square) and 8000 (triangle) grains.
For small vales of $s_s$, all curves fall
on one another. The humps at large sizes shift linearly with the average height of the heap.
(b) Distributions of the sizes of the avalanche clusters $(s)$
for the same systems as in top figure.
}}
\end{figure}


   The steady state of the system is characterised in two ways. The average area 
coverage over the whole granular system in the bin decreases with time
(here time is measured by the 
number of avalanches) and finally reaches a steady
value of 0.748 $\pm$ 0.005. A similar study but with different initial
configurations with a different value of average initial area
coverage shows that in the final steady state area coverage reaches the same value.
This implies that the final stable state is independent of the initial state
which is a requirement of Self-Organized Criticality.

   However, the area coverage has a local variation also and
in the steady state it is found to vary with the vertical height, the granular
pattern becoming denser with increasing height. Most of the
arches reside at the bottom creating more vacant regions at the bottom.
There are many avalanches which do not reach the surface, therefore in the
upper part of the pattern not many arches are formed and packing is more 
compact. We expect that the area coverage should grow to a maximum value of 
0.8180 $\pm$ 0.0002 near the top of the pile
as quoted in \cite{meakin} for the initial configuration
at $t=0$.

  The steady state may also be characterised by the distribution of the angles of the
contact vectors. When two grains are in contact, we draw two vectors
from their centres to the contact points. The angle $\phi$ is the angle
between the contact vector and the $+x$ direction. We measured the
probability distributions $P(\phi)$ in the range $0^o-360^o$ for the
initial state as well as in the steady state. To arrive at the steady
state we discard $2N$ avalanches, though the steady state attains   
even earlier than that.

  The contact angle distributions are shown in Fig. 9. The plot of
$P(\phi)$ vs. $\phi$ for the initial grain distribution shows four sharp peaks at
$60^o, 120^o, 240^o, 300^o$. We explain them in the following way.
If the grains were placed at the bottom in a complete orderly way by
mutually touching one another without any gap, the structure generated
by the BDRM method would be the hexagonal close packing (HCP) structure where
every grain has six other grains in contact at the interval of $60^o$s.
However, in our case, though we dropped grains randomly at the bottom level,
the deterministic piling process retains some effect of the HCP structure.
On the other hand, it is also known that the probability of a grain
having the number of contact neighbours different from four, decreases with the height as a
power law \cite {meakin}, which is certainly consistent with the four peaks we obtained.
There are also two small peaks at $0^o$ and at $180^o$ due
to the horizontal contacts of grains near the base level. This
distribution, however, changes drastically by reducing the heights of the peaks and
by broadening their widths in the steady state.
The randomness introduced into the system by randomly selecting
grains at the bottom increases the possibilities of other contact angles
also, but is not fully capable to totally randomize it to a flat distribution.
We check that after a large number of avalanches, this steady state distribution
remains unchanged.

  How big are the displacements of the individual grains taking part
in the avalanche? The
displacement $\Delta s$ is the absolute magnitude of the displacement
of a grain before and after an avalanche. We observe that there is a huge
variation the displacements $\Delta s$ of the grains. Most of the grains
displace very little, whereas others have much bigger displacements, but
their numbers are small. The displacement distribution $P(\Delta s)$ is
measured in the steady state over a large number of avalanches.
The lower cut-off of the distribution turned out to be strongly
dependent on the tolerance factor $\epsilon_1$ used for the simulation.
The upper cut-off of the distribution is of the order of unity since
when a grain is deleted at the bottom, its neighbouring grains drops
a distance of the order of the grain diameter.

  Systems of four different sizes have been simulated with $N$ = 300, 625,
2500 and 10000 in bins with base sizes $L_x$ = 20, 40, 80 and 160
units respectively. First $5N$ avalanches are thrown away
to allow the system to reach the steady state. A list is maintained
for the coordinates of all grains. This list is compared before and
after the avalanche to measure $\Delta s$. We observe a nice straight
line curve over many decades of the
$P(\Delta s)$ vs. $\Delta s$ plot on a double logarithmic scale, with
little curvature at the right edge implying 
a power law distribution for the displacement distribution(Fig. 10).
However, the slope of this curve
is found to increase with the system size and on extrapolating the slopes
with $L^{-1/2}$ we obtain the asymptotic slope ($N \rightarrow \infty$ limit) to be 0.97 $\pm$ 0.05.
We get similar behaviours for the distributions of the $x$ and $y$
components of the displacement vectors as well.

       In a stable configuration, the grains residing at the bottom
supports the weights of the grains above. A grain at the
bottom supports the partial weights of the few other neighbouring
contact grains residing above it. These neighbouring grains also
supports further neighbouring grains. Therefore, each grain at the
bottom supports a large number of grains above it. We call the set
of supporting grains as the `supporting cluster' corresponding to the
grain at the bottom (Fig. 11).

       We measure the size of the supporting cluster $s_s$ as the
number of supported grains. While creating an avalanche,
we randomly select a grain at the bottom. Before deleting it, we
calculate the value of $(s_s)$. The supporting cluster distribution
is plotted using a double logarithmic scale
in Fig. 12(a) for three system sizes of 2000, 4000 and 8000
grains contained within a bin of size $L_x$ = 100. We see that each
plot is having two distinctly separate regions. For small values
of $s_s$ we see a sharply decreasing variation, far from being
a power law. For large values of
$s_s$ we see a peaked variation. We explain, that in most cases,
the supporting cluster reaches the surface of the heap. However, in
some cases, the supporting cluster is completely surrounded by
an arch and therefore the cluster cannot be extended to the surface.
Since in the steady state, the arches are not of all length scales,
and has a characteristic size, the supporting clusters which do not
reach the surface also has a characteristic size.

       The avalanche cluster size $s$ is measured by the total number of
grains displaced in an avalanche in comparison with the initial configuration.
A snapshot of the configuration showing the supporting cluster, avalanche
cluster as well as the set of undisturbed grains are shown in Fig. 11.
The avalanche cluster size distribution is also plotted
using a double logarithmic scale in Fig. 12(b). For small values of $s$ we
get a value of the slope around 2.95. If we assume that the distribution
is described by a power law for small $s$, as $D(s) \sim s^{-\tau}$, then
$\tau = 2.95$.

  Now we comment on the effect of the parameter $\delta$. Since, during
the avalanche, different grains are updated in a sequence, the final steady
configuration depends on the sequence in which different grains were
updated. In the limit of $\delta \rightarrow 0$, this dependence will vanish.
We observe that, on decreasing $\delta$, the avalanche cluster size increases and
in the limit of $\delta \rightarrow 0$, the avalanche cluster and the supporting
cluster should be the same.

\begin{figure}
\begin{center}
\centerline{\epsffile{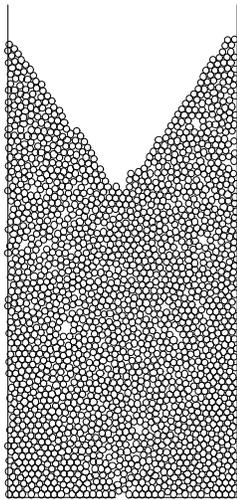}}
\narrowtext{\caption{
A granular system of 2000 grains of radii $1/\sqrt 2$ unit in a bin of width
50 units. The bin has a hole of width 3 units at the centre. The figure
shows the steady state of the system after 3000 avalanches.
}}
\end{center}
\end{figure}
\vskip -1.0 cm

  Therefore, we believe that the exponent $\tau$=2.95 is actually
a result of the finite $\delta = r/10$ used in our simulation. If we reduce
delta with the $D(s)$ vs. $s$ plot should look alike as the $D(s_s)$ vs. $s_s$
plot. To show that numerically, however, turned out to be difficult. We simulated a
particular cluster in a system of $N=2500, L_x=80, s_s=2068$, using $\delta, 10^{-1}\delta,
10^{-2}\delta $. The avalanche cluster sizes obtained are: 1046, 1080, 1081 respectively.

  We conclude that, small avalanches, which do not reach the granular surface
are unlikely to be of all sizes because the lengths of the arches present in the
system in the stationary state are not of all sizes. Therefore an arbitrary avalanche
is most likely to reach the surface. However, the avalanche size has a distribution,
which looks like a power law for finite $\delta$ but we expect that this behaviour is
temporary and in the limit of $\delta \rightarrow 0$, the distribution should fall
faster like that of the supporting cluster size distribution.

\vskip 0.2 cm
\leftline {\bf 5.2. Internal Avalanches of the Second Kind}
\vskip 0.2 cm

The flow of granular mass through a hole at the bottom of a granular bin
is studied here. If the hole is sufficiently small, the outflow of grains stops soon due to the
formation of a stable arch of grains clogging the hole. If the arch is broken,
the grains in the arch become unstable and a cascade of grain displacements
propagates within the system what we call as the
Internal Avalanche of the second kind. As a result,
a fresh flow starts through the hole,
which also eventually stops due to the formation of another arch. It is known that
the amount of out flowing granular mass between successive clogging events varies
over a wide range, however, on average it depends sensitively on the hole
diameter as compared to the grain size.

We study distribution of outflow sizes from a granular bin in two-dimensions and 
observe its power law variation \cite {clogging}.

\begin{figure}
\begin{center}
\epsfxsize=3.3in
\epsffile{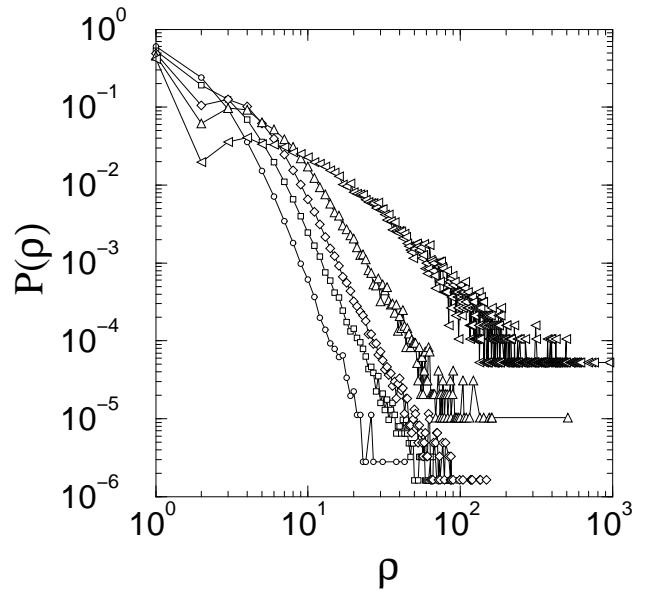}
\narrowtext{\caption{
The distribution $P(\rho)$ of the number of grains $\rho$ that flows out of the system in an
internal avalanche of the second kind between two successive clogged situations
plotted on a double logarithmic scale. A bin of size $Lx=50$, containing
4000 grains of radii $r=1/\sqrt 2$ was simulated with the hole sizes $R$ = 0.75, 1, 1.25, 1.5, and 2.0
(curves from left to right).
}}
\end{center}
\end{figure}
\vskip -0.5 cm

We study a two-dimensional granular system with grains kept in a vertical
rectangular bin. Initially the system has no hole at the bottom and we
fill the bin up to a certain height. A hole of half-width $R$
is then made at the centre of the bottom. Some grains drop out through the hole. This
flow is stopped when an `arch' is formed
which covers the hole, forbids other
grains to flow out and the system attains a stable state. 
The two lower most grains of the arch on the two sides of
the hole are chosen   and one of them is selected randomly.
By deleting this grain, the arch is made unstable, which creates an internal
avalanche of grain movements within the bin.
Grains which come down and touch the bottom
line of the bin with their centres within the hole are
removed. While removing such a grain, the whole avalanche
dynamics is frozen and the grain is replaced back on the top
surface again by the BDRM method. A granular pattern after a large number of
avalanches of second kind is shown in Fig. 13.

We count the number of grains $\rho$ that flow out of the system
during an avalanche between two successive clogging events and
study its distributions for various values of the hole sizes $R$.
We used a bin of dimensions $L_x=50$ and $L_y=200$ units containing
4000 grains of radii $r=1/\sqrt 2$.

In the steady state, two conical shapes are formed. The grains which are
never disturbed by any avalanche are located on both sides
of the hole. These undisturbed grains form a cone.
The second cone is formed on the upper surface, though the grains
were dropped on this surface along randomly chosen horizontal
coordinates with uniform probability. For a fixed size of the base,
the angle of the cone is found to increase with the
average height of the granular column.

Holes of five different
sizes, i.e., $R = 0.75, 1, 1.25, 1.5, 2$ are used. We could generate
around $5 \times 10^5$ such outflows for each hole size
and collected the distribution data for $P(\rho)$.
We show the plot of these curves in Fig. 14 on a double logarithmic scale.
The grain which is deleted to break the arch is also counted in $\rho$.
It is found increasingly improbable that the outflow will consist of
only another grain. Consequently $P(1)$ is found to decrease monotonically
on increasing $R$. However, for large values of $\rho$, $P(\rho)$ decays.

  All curves of $P(\rho)$ vs. $\rho$ plots show straight lines for large
$\rho$ values, signifying the power law distributions for the avalanche sizes.
The slopes of these curves $\sigma (R)$ are found to depend strongly on the
size of the hole, as well as the grain size: $P(\rho) \sim \rho^{-\sigma(R,r)}$.
We observe that $\sigma(R,r)$ is actually a linear function of $(R/r)$
as $\sigma(R,r)=a - k(\frac{R}{r})$ with $k$=1.6.
The average outflow $<\rho(R/r)>$ diverges exponentially as:
$exp(\alpha(\frac{R}{r}-1))$ with $\alpha = 0.7$.

Financial support by the Indo-French Centre for the Promotion of Advanced 
Research (IFCPAR) through the project No. 1508-3 is gratefully acknowledged. 
I also thank my collaborators in this project: Hans. J. Herrmann, Devang V. 
Khakhar, Stephane Roux and Supriya Krishnamurthy.

\vskip 0.2 cm

\leftline {Electronic Address: manna@fermion.bose.res.in}

\end{multicols}

\end {document}